\documentclass[a4paper,onecolumn,11pt,unpublished,accepted=2023-07-04]{quantumarticle}
\pdfoutput=1
\usepackage[utf8]{inputenc}
\usepackage[english]{babel}
\usepackage[T1]{fontenc}
\usepackage{amsmath}
\usepackage{hyperref}

\usepackage[numbers,compress]{natbib}

\usepackage{algorithm}
\usepackage[noend]{algpseudocode}

\usepackage{tikz}
\usepackage{lipsum}
\usepackage{bbold}

\usepackage{outlines}

\begin{document}

\title{Quantum computing Floquet energy spectra}

\author{Benedikt Fauseweh}
\email{benedikt.fauseweh@dlr.de}
\affiliation{Institute for Software Technology, German Aerospace Center (DLR),  Linder H\"ohe, 51147 Cologne, Germany}
\orcid{0000-0002-4861-7101}
\author{Jian-Xin Zhu}
\email{jxzhu@lanl.gov}
\affiliation{Theoretical Division, Los Alamos National Laboratory, Los Alamos, New Mexico 87545, USA}
\affiliation{Center for Integrated Nanotechnologies, Los Alamos National Laboratory, Los Alamos, New Mexico 87545, USA}
\maketitle

\begin{abstract}
Quantum systems can be dynamically controlled using time-periodic external fields, leading to the concept of Floquet engineering, with promising technological applications. Computing Floquet {energy spectra} is harder than only computing ground state properties or single time-dependent trajectories, and scales exponentially with the Hilbert space dimension. Especially for strongly correlated systems in the low frequency limit, classical approaches based on truncation break down. Here, we present two quantum algorithms to determine effective Floquet modes and {energy spectra}. We combine the defining properties of Floquet modes in time and frequency domains with the expressiveness of parametrized quantum circuits to overcome the limitations of classical approaches. We benchmark our algorithms and provide an analysis of the key properties relevant for near-term quantum hardware. 
\end{abstract}

\section{Introduction}

\begin{figure}[t]
  \centering
  \includegraphics[width=0.99\textwidth]{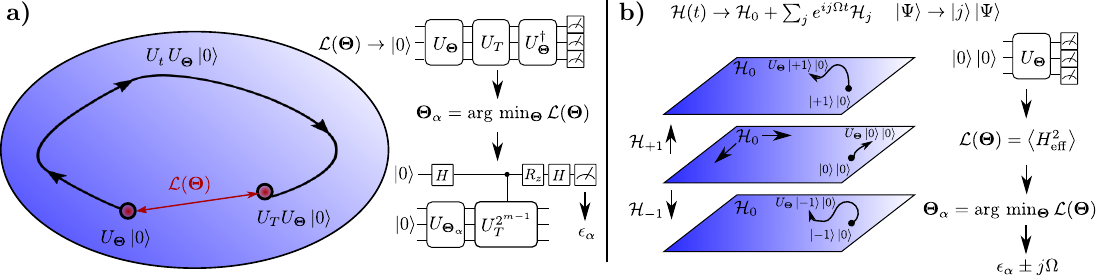}
  \caption{ Graphical representation of both algorithms for the determination of Floquet eigenstates and energies. \textbf{a)} Algorithm Fauseweh-Zhu-1. The parametrized ansatz circuit $U_\mathbf{\Theta} \left| 0 \right\rangle$ is time evolved over a full period and the overlap to its starting state is calculated. Once the overlap is maximized, we compute the Floquet quasi-energy using iterative quantum phase estimation. \textbf{b)} Algorithm Fauseweh-Zhu-2. The Hilbert space is extended by Fourier expansion of the Hamiltonian. A parametrized quantum circuit approximates the combined Floquet state described by both physical and Fourier quantum numbers. Excited state VQE is used to determine Floquet states and energies. }
  \label{fig:figure1}
\end{figure}

The interaction between the electromagnetic field and matter is one of the basic principles, which is used to probe and manipulate electronic and magnetic degrees of freedom. Quantum systems that are subjected to time-periodic irradiation show intriguing phenomena such as light-induced surface states \cite{doi:10.1126/science.1239834}, topological phases of matter \cite{PhysRevLett.111.185302, PhysRevLett.111.185301, Jotzu2014,Rechtsman2013,Weitenberg2021}, high harmonic generation \cite{Ghimire2019} and light-induced superconductivity \cite{doi:10.1080/00107514.2017.1406623}. With advances in high-power ultra-fast spectroscopy, non-linear phenomena, including multi-photon processes, have opened up new prospects of dynamically controlling quantum dynamics, engineering atomic \cite{RevModPhys.81.163}, molecular \cite{doi:10.1126/science.288.5467.824,doi:10.1126/science.1059133, PhysRevLett.114.233003} and solid-state \cite{doi:10.1126/science.1197294,RevModPhys.83.471} properties on demand.
On the theoretical side the study of non-equilibrium dynamics in highly entangled many-body systems is at the forefront of current research \cite{RevModPhys.83.863,RevModPhys.91.021001,doi:10.1146/annurev-conmatphys-031218-013721,Schwarz2020,PhysRevB.102.165128,PhysRevB.101.224510,PhysRevB.103.045132,PhysRevB.103.224305,PhysRevB.101.180507}.   
Light-driven quantum systems are described by time-dependent Hamiltonians $H(t)$, which include the interaction between the external electromagnetic field and the electronic or magnetic degrees of freedom. \\
If the field is periodic in time, i.e., $H(t) = H(t+T)$ for some period $T$, one can use the concept of Floquet engineering to control the microscopic degrees of freedom \cite{doi:10.1146/annurev-conmatphys-031218-013423}. Floquet engineering is based on the Floquet theorem \cite{PhysRev.138.B979,PhysRevA.7.2203}. It states that there exists a complete and orthogonal set of solutions $\left| \Psi (t) \right\rangle$ to the time-dependent Schrödinger equation
\begin{align}
i \frac{\mathrm{d}}{\mathrm{d}t} \left| \Psi (t) \right\rangle = H(t) \left| \Psi(t) \right\rangle,
\end{align}
taking the form
\begin{align}
\label{eq.floquetMode}
\left| \Psi_\alpha (t) \right\rangle = \left| \Phi_\alpha (t) \right\rangle e^{- i \epsilon_\alpha t},  \, \, \left| \Phi_\alpha (t+T) \right\rangle = \left| \Phi_\alpha (t) \right\rangle \;.
\end{align}
Here we have chosen the reduced Planck constant $\hbar=1$, $\vert \Phi_{\alpha}(t)\rangle$ is the Bloch amplitude periodic in time, and $\epsilon_\alpha$ denotes the Floquet \textit{quasi-energy}. The quasi-energies depend on the eigenstates of the non-driven system as well as the specific form of the driving field. They are uniquely defined up to the fundamental frequency $\Omega = 2 \pi / T$. Changing the modulation of the driving field, such as shape, intensity and frequency allows to modify the $\epsilon_\alpha$'s and thereby the quasi-energy or Floquet {\textit{energy spectrum}}, inducing novel topological phases not present in equilibrium \cite{PhysRevX.6.041001}. \\
While the Floquet theorem is a strong statement, computing the quasi-energy {spectrum} explicitly, depending on the microscopic degrees of freedom and the time-dependence of the external field, remains a difficult problem due to the exponentially scaling Hilbert space. Various classical techniques, such as  time-dependent dynamical mean-field theory (t-DMFT) \cite{PhysRevB.78.235124,PhysRevLett.103.047403,PhysRevB.89.205126,PhysRevB.96.045125, PhysRevB.96.075134}, time-dependent density matrix renormalization group (t-DMRG) \cite{PhysRevB.82.205110}, kinetic equations \cite{PhysRevX.5.041050,PhysRevB.90.195429}, perturbative high-frequency expansions \cite{doi:10.1080/00018732.2015.1055918,Mentink2015,RevModPhys.89.011004,PhysRevB.95.014112,PhysRevLett.101.245302,Eckardt_2015,Rodriguez_Vega_2018} and exact diagonalization \cite{PhysRevB.66.205320,PhysRevResearch.4.043174,PhysRevB.93.155132} have been employed, but unfortunately most of them either are not universally applicable or scale exponentially in system size. Especially the computation of the whole quasi-energy { spectrum} requires more than a simple time-evolution. \\
Another problem in Floquet-engineering comes from the limited theoretical modeling, in which only a few band model is considered and the perturbative infinite frequency expansion can be applied. In real materials however the situation is more difficult, as higher energy bands become important once the frequency is sufficiently large to induce optical transitions.\\
In this work, we propose to use quantum computers to overcome the limitations of previous classical approaches. We present two algorithms that in principle can calculate all Floquet eigenstates and quasi-energies. Our quantum-classical hybrid approach takes the limitations of modern Noisy Intermediate-Scale Quantum (NISQ) hardware \cite{Preskill2018quantumcomputingin} into account and allows to either improve accuracy by increasing the number of auxiliary qubits or by deepening the quantum circuits. Both algorithms make use of the enhanced expressiveness \cite{PhysRevResearch.2.033125} of parametrized quantum circuits to find approximate Floquet eigenstates and compute the corresponding quasi-energies. While the first algorithm works in the original Hilbert space of the time-dependent problem, the second algorithm makes use of the extended Floquet Hilbert space in frequency domain. After a detailed explanation of the basic principles, we evaluate the performance of both algorithms by investigating the simplest system that is not analytically solvable: a spin-$\frac{1}{2}$ in a linearly polarized periodic magnetic field.

\section{First Algorithm}

In the first algorithm, \textit{Fauseweh-Zhu-1}, we use a parametrized quantum circuit $U_{\mathbf{\Theta}}$, with variational parameters $\mathbf{\Theta}$, in combination with the defining properties of the time evolution operator 
\begin{align}
\left(H(t) - i \frac{\mathrm{d}}{\mathrm{d}t} \right) U(t, t_0) = 0, \quad U(t_0, t_0) = \mathbb{1},
\end{align}
to determine a good approximation for Floquet eigenstates. Without loss of generality, we set $t_0 = 0$ and denote $U_t =  U(t, 0)$. The Floquet theorem implies that for a full time period $t=T$ the time evolution operator has complex eigenvalues of the form
\begin{align}
U_T \left| \Psi_\alpha (0) \right\rangle = \left| \Psi_\alpha (T) \right\rangle = e^{-i \epsilon_\alpha T} \left| \Psi_\alpha (0) \right\rangle 
\end{align}
We represent the Floquet eigenstate $\left| \Psi_\alpha (0) \right\rangle$ using a parametrized quantum circuit
\begin{align}
\left| \Psi_\alpha (0) \right\rangle \approx U_{\mathbf{\Theta}}
 \left| 0 \right\rangle,
\end{align}
where $ \left| 0 \right\rangle$ is the initial state of the quantum computer. If $U_{\mathbf{\Theta}} \left| 0 \right\rangle$ is a Floquet eigenstate, then 
\begin{align}
\left|  \left\langle 0 \right| U_{\mathbf{\Theta}}^\dagger   U_T U_{\mathbf{\Theta}}
  \left| 0 \right\rangle \right|^2 = 1
\end{align}
holds for the overlap between the time-evolved state and the initial state. Note that overlaps of the form $ | \langle 0 | U^\dagger V | 0 \rangle  |^2$ can be efficiently computed on a quantum computer by interpreting them as the measurement of the computational initial state projector
\begin{align}
| \langle 0 | U^\dagger V | 0 \rangle  |^2 = \langle 0 |   V^\dagger U    | 0  \rangle  \langle 0  |   U^\dagger V   | 0  \rangle .
\end{align}
Thus we devise a hybrid algorithm that maximizes the overlap to determine a Floquet eigenstate. An optimal solution $\mathbf{\Theta}_\alpha$ has a maximal overlap of $1$. 
To obtain the complete set of eigenstates, we modify the target function using a Lagrange multiplier $\lambda > 0$ to project out solutions $\mathbf{\Theta}_{\beta}$ that have previously been computed:
\begin{align}
\mathcal{L}({\mathbf{\Theta}}) = \left|  \left\langle 0 \right| U_{\mathbf{\Theta}}^\dagger    U_T U_{\mathbf{\Theta}}
  \left| 0 \right\rangle \right|^2 - \lambda \sum\limits_{\beta} \left|  \left\langle 0 \right| U_{\mathbf{\Theta}_\beta}^\dagger    U_{\mathbf{\Theta}}
  \left| 0 \right\rangle \right|^2.
\end{align}
Here $\lambda$ has to be choosen sufficiently large for the algorithm to find a new solution, see also \cite{Higgott2019variationalquantum}.
Once a Floquet eigenstate has been determined, we use iterative quantum phase estimation (IQPE) \cite{PhysRevLett.76.3228,PhysRevLett.98.090501,PhysRevA.76.030306} to compute the complex phase $e^{-i \epsilon_\alpha T}$ of $U_T$ upon application to the eigenstate. The algorithm is sketched in Fig. \ref{fig:figure1} a). 

\begin{algorithm}
\caption{Fauseweh-Zhu-1}\label{algo1}
\begin{algorithmic}[1]
\Require 
\Statex parametrized quantum circuit $U_{\mathbf{\Theta}}$
\Statex previous solutions $\mathbf{\Theta}_\beta$
\Statex periodic Hamiltonian $\mathcal{H}(t)$
\Procedure{Optimize}{$U_{\mathbf{\Theta}}$}
\State Choose initial parameters $\mathbf{\Theta}$
\While{Maximize $\mathcal{L}(\mathbf{\Theta}) = \left|  \left\langle 0 \right| U_{\mathbf{\Theta}}^\dagger  \left.  \right|  U_T^{\phantom\dagger} U_{\mathbf{\Theta}}^{\phantom\dagger}
  \left| 0 \right\rangle \right|^2 - \lambda \sum_{\beta} \left|  \left\langle 0 \right| U_{\mathbf{\Theta}_\beta}^\dagger   \left.  \right|  U_{\mathbf{\Theta}}
  \left| 0 \right\rangle \right|^2$ }
  	\State Evaluate circuit $ U_{\mathbf{\Theta}}^\dagger U_T U_{\mathbf{\Theta}}^{\phantom\dagger} 
  \left| 0 \right\rangle$
    \State Evaluate circuits $ U_{\mathbf{\Theta}_\beta}^\dagger U_{\mathbf{\Theta}}^{\phantom\dagger} 
  \left| 0 \right\rangle$
  	\State Update parameters $\mathbf{\Theta}$ to increase target $\mathcal{L}(\mathbf{\Theta})$ 
\EndWhile
\State Iterative quantum phase estimation on optimized state $U_\mathbf{\Theta_\alpha} \left| 0 \right\rangle$ for $\epsilon_\alpha$
\State \Return{$\epsilon_\alpha,\, \mathbf{\Theta_\alpha}$}
\EndProcedure
\end{algorithmic}
\end{algorithm}

We turn to a high-level analysis of the algorithm and its requirements. The basis for the algorithm is the optimization of a parametrized quantum circuit. Thus one of the fundamental requirements is that the manifold spanned by all the possible quantum circuits within the ansatz \cite{Funcke2021dimensional} contains the Floquet eigenstates or is at least close to it. Thus carefully choosing a circuit ansatz is important for the applicability of our algorithm, and is currently a subject of intensive research \cite{Kandala2017,PhysRevA.92.042303,PhysRevResearch.3.023092,Gard2020,Cerezo2021,PhysRevApplied.11.044092}. Note that the target function $\mathcal{L}({\mathbf{\Theta}})$ allows for a gradient-based optimization using parameter shift rules \cite{PhysRevLett.118.150503}. 

The algorithm optimizes a global observable and is therefore, in principle subject to the problem of barren plateaus \cite{Cerezo2021_2}. This can be avoided by replacing the global observable with a local observable with identical extremum, see also \cite{Barison2021efficientquantum}.

We do not specify how the time evolution in the first part of the algorithm is performed. Trotterization is in principle applicable to $k$-local Hamiltonians \cite{doi:10.1126/science.273.5278.1073} and is the most accurate. However,  it is disadvantageous with respect to circuit depth \cite{Fauseweh2021,Smith2019,PhysRevLett.121.170501}. Hence it is best applied if $T$ is the smallest time scale, i.e., in the large frequency limit. Various other methods have been developed \cite{PhysRevLett.114.090502,PhysRevLett.123.070503,Childs2019fasterquantum,Low2019hamiltonian,Cirstoiu2020} that make use of shallower circuits, including approaches that work directly within the variational manifold \cite{Yuan2019theoryofvariational,Barison2021efficientquantum,otten2019noiseresilient}. All of these methods can be combined with our algorithm, as long as they keep track of the global phase alignment. 

The final part of the algorithm uses a single ancillary qubit to perform IQPE. It applies controlled $(U_T)^{2^n}$ gates to compute the $n$-th bit of the phase $2\pi \phi = \epsilon_\alpha T $. The iteration depths $n_\mathrm{max}$ of IQPE determines the phase resolution and thereby desired precision of the quasi-energies. Unless powers of $U_T^2$ can be obtained within the same circuit depth, e.g., with variational time evolution, IQPE leads to an increasing circuit depth requirement for increasing precision.

\section{Second algorithm}

The second algorithm, \textit{Fauseweh-Zhu-2}, uses Fourier analysis to map the problem onto an extended Hilbert space that is then solved using variational quantum eigensolver approaches.
From the definition of the Floquet modes in \eqref{eq.floquetMode} one can derive that the Hermitian operator $\mathcal{H}(t) =  H(t) - i \frac{\mathrm{d}}{\mathrm{d}t}$ has eigenvalues 
\begin{align}
\mathcal{H}(t) \left| \Phi_\alpha (t) \right\rangle = \epsilon_\alpha \left| \Phi_\alpha (t) \right\rangle.
\end{align}
Applying a Fourier expansion we can map the time-dependent evolution problem to a time-independent eigenvalue problem
\begin{align}
H(t) = \sum\limits_{j} e^{-i j \Omega t} H_j , \quad  \left| \Phi_\alpha (t) \right\rangle = \sum\limits_j e^{-i j \Omega t} \left| \Phi_\alpha^j \right\rangle \\
\Rightarrow \sum\limits_{j} \left( H_{j-k} - j \Omega \delta_{j,k} \right) \left| \Phi_\alpha^j \right\rangle = \epsilon_\alpha \left| \Phi_\alpha^k \right\rangle . \label{eq.Eigenvalue}
\end{align}
Note that the state $\left| \Phi_\alpha^k \right\rangle$ is part of an extended Hilbert space $\mathcal{R} \otimes \mathcal{T}$ containing the original Hilbert space $\mathcal{R}$ onto which $H(t)$ acts and the Hilbert space $\mathcal{T}$ of square-integrable periodic functions of period $T$. This extended Hilbert space is not physical, but it has a clear connection to the original Hilbert space: the eigenvalues of the time-independent problem are identical to the original time-dependent problem, up to multiples of $j \Omega$. A graphical interpretation of this procedure is shown in Fig. 1 \textbf{b)}, as the original system is expanded into infinitely many copies. The Hamiltonian $H_0$ acts within the plane, while $H_{\pm \gamma}, \gamma > 0$ introduces hopping in the $j$ direction. In the literature, this is understood as extending the system in a new dimension \cite{doi:10.1146/annurev-conmatphys-031218-013423}. This is certainly true for non-interacting systems, but not in the strict many-body sense, as with each additional layer in the quantum number $j$ the Hilbert space dimension increases by dim($\mathcal{R}$), but in a many-body system it would be multiplied by dim($\mathcal{R}$).

The eigenvalue problem in \eqref{eq.Eigenvalue} is the starting point for our second algorithm. We parameterize the Floquet eigenstates $\left| \Phi_\alpha^k \right\rangle$ using a parametrized quantum circuit
\begin{align}
\left| \Phi_\alpha^k \right\rangle \approx U_{\mathbf{\Theta}} \left| 0 \right\rangle_{\mathcal{T}} \left| 0 \right\rangle_{\mathcal{R}} .
\end{align}
Notice that we used the notation $\left| 0 \right\rangle _{\mathcal{T}} \left| 0 \right\rangle_{\mathcal{R}}$ to mark the extended Hilbert space computational basis. Here the first number marks the quantum number $j$ of the $\mathcal{T}$ Hilbert space part, while the second number refers to the original Hilbert space $\mathcal{R}$. In the following we neglect the indices $\mathcal{T}$ and $\mathcal{R}$ of the states. We define the effective Hamiltonian
\begin{align}
H_\mathrm{eff}^{j,k} = H_{j-k} - j \Omega \delta_{j,k}.
\end{align}
Now we want to compute the eigenstates of $H_\mathrm{eff}$ to obtain the Floquet {energy spectrum}. We use the variational principle to optimize the parameters $\mathbf{\Theta}$. As the extended Hilbert space is infinite dimensional, we truncate $H_\mathrm{eff}$ at $\pm j_\mathrm{max}$. This truncation introduces finite size errors in our calculation. We expect that finite size errors are smallest in the center of the {energy spectrum}. We therefore minimize $H_\mathrm{eff}^2$ instead of $H_\mathrm{eff}$, giving us the squared eigenvalues $\epsilon_\alpha^2$. The sign of the eigenvalues can then be determined after optimization by measuring the expectation value of $H_\mathrm{eff}$. As in the first algorithm we define our target function using a Lagrange multiplier $\lambda > 0$ to project out solutions $\mathbf{\Theta}_{\beta}$ that have previously been computed
\begin{align}
\mathcal{L}(\mathbf{\Theta}) =  \left\langle 0 \right| \left\langle 0 \right| U_{\mathbf{\Theta}}^\dagger H_\mathrm{eff}^2 U_{\mathbf{\Theta}}^{\phantom\dagger}
  \left| 0 \right\rangle \left| 0 \right\rangle - \lambda \sum_{\beta} \left|  \left\langle 0 \right| \left\langle 0 \right| U_{\mathbf{\Theta}_\beta}^\dagger    U_{\mathbf{\Theta}}
  \left| 0 \right\rangle \left| 0 \right\rangle \right|^2,
\end{align}
which corresponds to an excited state variational quantum eigensolver \cite{Higgott2019variationalquantum} for the squared effective Hamiltonian. An overview of the Algorithm is sketched in Fig. 1 \textbf{b)}. 

\begin{algorithm}
\caption{Fauseweh-Zhu-2}\label{algo2}
\begin{algorithmic}[1]
\Require 
\Statex parametrized quantum circuit $U_{\mathbf{\Theta}}$ in extended Hilbert space $\mathcal{R} \otimes \mathcal{T}$
\Statex previous solutions $\mathbf{\Theta}_\beta$
\Statex Fourier expansion of Hamiltonian $H_\mathrm{eff} = H_{j-k} - j \Omega \delta_{j,k}$
\Statex Truncation value $j_\mathrm{max}$
\Procedure{Optimize}{$U_{\mathbf{\Theta}}$}
\State Choose initial parameters $\mathbf{\Theta}$
\While{Maximize $\mathcal{L}(\mathbf{\Theta}) =  \left\langle 0 \right| \left\langle 0 \right| U_{\mathbf{\Theta}}^\dagger H_\mathrm{eff}^2 U_{\mathbf{\Theta}}^{\phantom\dagger}
  \left| 0 \right\rangle \left| 0 \right\rangle - \lambda \sum_{\beta} \left|  \left\langle 0 \right| \left\langle 0 \right| U_{\mathbf{\Theta}_\beta}^\dagger   \left.  \right|  U_{\mathbf{\Theta}}
  \left| 0 \right\rangle \left| 0 \right\rangle \right|^2$ }
  	  \State Evaluate observable $\left\langle 0 \right| \left\langle 0 \right| U_{\mathbf{\Theta}}^\dagger H_\mathrm{eff}^2 U_{\mathbf{\Theta}}^{\phantom\dagger}
  \left| 0 \right\rangle \left| 0 \right\rangle$
      \State Evaluate circuits $ U_{\mathbf{\Theta}_\beta}^\dagger U_{\mathbf{\Theta}}^{\phantom\dagger} 
  \left| 0 \right\rangle \left| 0 \right\rangle$
  	\State Update parameters $\mathbf{\Theta}$ to increase target $\mathcal{L}(\mathbf{\Theta})$ 
\EndWhile
\State Compute $\epsilon_\alpha \pm j \Omega = \left\langle 0 \right| \left\langle 0 \right| U_{\mathbf{\Theta}}^\dagger H_\mathrm{eff} U_{\mathbf{\Theta}}^{\phantom\dagger}
  \left| 0 \right\rangle \left| 0 \right\rangle$
\State \Return{$\epsilon_\alpha \pm j \Omega,\, \mathbf{\Theta_\alpha}$}
\EndProcedure
\end{algorithmic}
\end{algorithm}

By analyzing the second algorithm in comparison to the first algorithm, we immediately identify the increased auxiliary qubits that are required due to the extension of the Hilbert space. While this increases the width of the quantum circuit, its depth now depends purely on the depth of the ansatz. This is in striking contrast to the first algorithm, where the maximum depth is determined by the IQPE part of the algorithm, and hence by the required quasi-energy resolution. 
The second algorithm also has the property of benefiting from a mixed qudit-qubit architecture, as the truncated $ \mathcal{T}$ part of the Hilbert space naturally leads to states such as $\left| \pm j \right\rangle \left| \phi \right\rangle$, with $\left| \phi \right\rangle$ being a  state in the $\mathcal{R}$ Hilbert space. This could be useful for quantum computer architectures that have access to more than two states per fundamental quantum building block. 
The accuracy of the approach also directly depends on the maximum truncation $j_\mathrm{max}$, that needs to be determined, depending on the driving and the original Hamiltonian.

\section{Benchmark}

{In the following we evaluate the applicability, performance and scalability of both algorithms. We start with a} simple system, that is not analytically solvable, the linearly driven spin-$\frac{1}{2}$,
\begin{align}
\label{eq.driven12}
H(t) = - \frac{\Delta}{2} \sigma_z + \frac{A}{2} \cos(\Omega t) \sigma_x \;.
\end{align}
Here $\sigma_{i}$ are the Pauli matrices. We fix the energy spacing $\Delta = 1$ and the driving frequency $\Omega = 2.5$, and investigate the Floquet {energy spectrum} as a function of the amplitude $A$ of the external field.
{To investigate how the approach scales for larger systems with strong correlations we then perform simulations of the Floquet states for the circularly driven spin-$\frac{1}{2}$ Heisenberg chain with periodic boundary conditions
\begin{align}
\label{eq.driven_chain}
H(t) = - \frac{J}{4} \sum\limits_{i} \sum\limits_{\alpha \in \lbrace x,y,z \rbrace} \sigma_{i, \alpha} \sigma_{i+1, \alpha}  + \sum\limits_{i } \left[  A \cos(\Omega t) \sigma_{i, x} +  A \sin(\Omega t) \sigma_{i, y} \right]  \;.
\end{align}
In equilibrium the spin-$\frac{1}{2}$ chain is gapless and exhibits fractionalized spinon excitations  \cite{FADDEEV1981375} in the thermodynamic limit. In a finite size systems elementary excitations are best described by gapped $S=1$ triplet excitations. 
%The spin-$\frac{1}{2}$ model 
%%
%What has been  done on Floquet states in Heisenberg chains??
%%
%Properties of the system ... ground state, triplon exitations, scaling of A with system size so that we are not in the random state limit
%Analysis of scaling -> Much not known about the properties of floquet states in Hilbert space, therefore analysis of depth of ansatz w.r.t. to precision
}

\subsection{Linearly driven spin-$\frac{1}{2}$}

\subsubsection{Algorithm Fauseweh-Zhu-1}

\begin{figure}[t]
  \centering
 \includegraphics[width=0.99\textwidth]{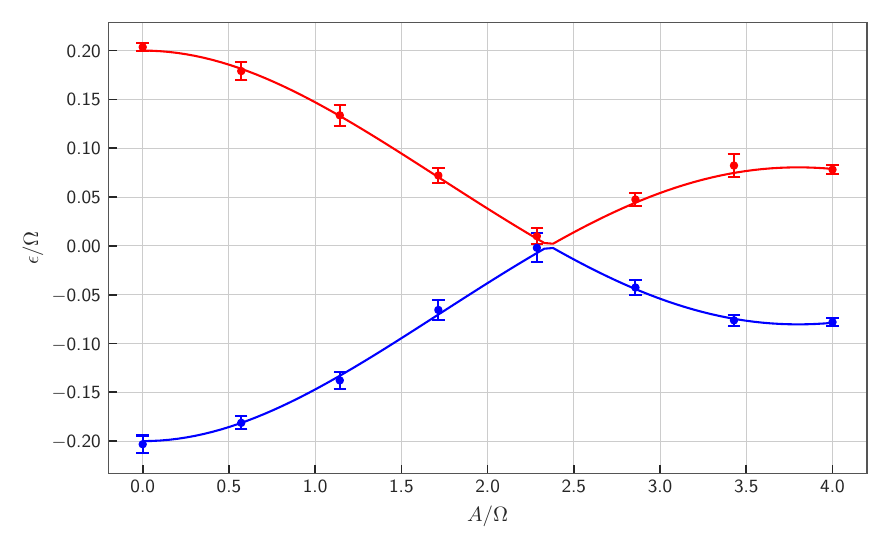}
  \caption{ Simulation results for the Floquet energies of the driven spin-$\frac{1}{2}$ system using algorithm Fauseweh-Zhu-1 with $\lambda = 5$. Straight lines show the exact result. Errorbars from sampling show the circular standard-deviation. }
  \label{fig:figure2}
\end{figure}

We use the $U3$ gate as a generic single qubit rotation for the parametrized quantum circuit:
\begin{align}
U3(\theta, \phi, \nu) =
    \begin{pmatrix}
        \cos(\frac{\theta}{2})          & -e^{i\nu}\sin(\frac{\theta}{2}) \\
        e^{i\phi}\sin(\frac{\theta}{2}) & e^{i(\phi+\nu)}\cos(\frac{\theta}{2})
    \end{pmatrix}
\end{align}
The time evolution $U_T = \mathcal{T} \exp(- i \int \mathrm{d}t  H(t))$ is performed using Trotterization with $100$ time steps. Optimization is performed using conjugate gradient descent on a quantum computer simulator using the IBM qiskit package \cite{Qiskit-Textbook}. The eigenstates are optimized with $10^4$ samples, while the IQPE uses $100$ samples with $5$ iterations. Results for the Floquet {energy spectrum} are shown in Fig.\ \ref{fig:figure2}. We see a very good agreement between the simulated results and the exact quasi-energy {spectrum}. The error bars are the result of the limited iteration depth in the IQPE. 

\subsubsection{Algorithm Fauseweh-Zhu-2}

\begin{figure}[t]
  \centering
  \includegraphics[width=0.99\textwidth]{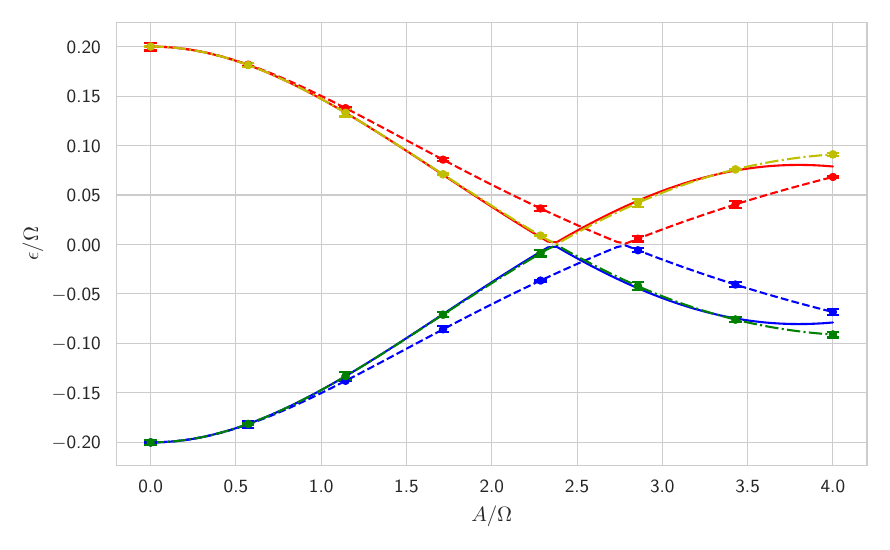}
  \caption{ Simulation results for the Floquet energies of the driven spin-$\frac{1}{2}$ system using algorithm Fauseweh-Zhu-2 with $\lambda = 5$. Lines show the exact result, dashed lines show the exact eigenstates of the truncated Hamiltonian {for   $j_\mathrm{max} = \pm 1$ and dashed dotted lines for  $j_\mathrm{max} = \pm 2$} in the extended Hilbert space. {Red and blue dots are for   $j_\mathrm{max} = \pm 1$ and green and yellow dots for $j_\mathrm{max} = \pm 2$}. Errorbars are from $10^4$ samples.}
  \label{fig:figure3}
\end{figure}

As the second algorithm uses the extended Hilbert basis we transfer the problem in \eqref{eq.driven12} to frequency space with a truncation value of $j_\mathrm{max} = \pm 1$
\begin{align}
H_\mathrm{eff} = \frac{\Delta}{2} \sigma_z \otimes \mathbb{1}  + \frac{A}{2} \sigma_x \otimes S_x + \Omega  \mathbb{1} \otimes S_z \;,
\end{align}
where the matrix $S_x$ is defined in the appendix \ref{appendix}. To account for this, we must also enlarge the ansatz in the parametrized quantum circuit and 
cannot use a single qubit gate. We use a variational Hamiltonian approach \cite{PhysRevA.92.042303} to effectively reach the target states,
\begin{align} \label{eq.algo2_ansatz_general}
U_{\mathbf{\Theta}} = \prod\limits_{l} \exp( - i K_l \Theta_l).
\end{align}
We specify the hermitian generators $K_l$ in appendix \ref{appendix}. The parameters are optimized using the conjugate gradient descent method. We used $10^4$ samples for optimization and evaluation of observables on a quantum computing simulator. For comparison we also computed the exact eigenvalues of the Floquet matrix. The results are shown in Fig.\ \ref{fig:figure3}. We see a very good agreement between the simulated results and the exact quasi-energy structure of the truncated Hamiltonian. Naturally the truncation leads to an increasing error with amplitude $A$ and only qualitatively captures the exact quasi-energy {spectrum} to a certain point.  {Increasing the truncation from $j_\mathrm{max} = \pm 1$  to $j_\mathrm{max} = \pm 2$ significantly extends the agreement range of the second algorithm with the exact result of the non-truncated Hamiltonian.}

\subsection{Circularly driven spin-$\frac{1}{2}$ Heisenberg chain}

{To evaluate the performance of our algorithms for the spin-$\frac{1}{2}$ chain we fix the Heisenberg interaction to $J=1$ and the field amplitude $A=2$. We perform simulation without shot or other noise on a quantum computer simulator. To evaluate the effect of hardware noise we perform simulations assuming a symmetric depolarizing noise channel with details given in Appendix \ref{appendix.b}.  As we are interested in Floquet states that still have a well defined energy spectrum, as opposed to essentially random states \cite{PhysRevX.4.041048}, we choose $\Omega = 5 J N_{\text{site}}$, where $N_\text{site}$ is the number of sites in the chain. This scaling of the driving frequency with the number of sites avoids that $U(T)$  exhibits properties of random matrices belonging to circular ensembles. Of course physically interesting cases in the thermodynamic limit do not necessarily fulfill this criterion, but there are numerical indications that in many cases there exists a critical frequency $\Omega_\mathrm{crit}$  separating regimes of finite and infinite heating \cite{PhysRevLett.80.1808,PhysRevE.60.3949,DALESSIO201319,CITRO2015694}, so that Floquet spectra are well defined even in the thermodynamic limit.  }

\subsubsection{Algorithm Fauseweh-Zhu-1}

\begin{figure}[t]
  \centering
  \includegraphics[width=0.99\textwidth]{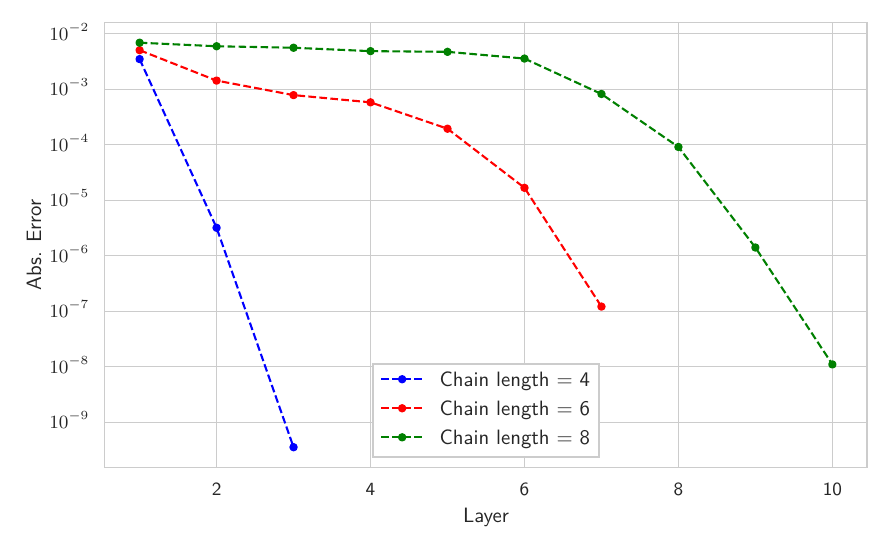}
  \caption{ {Simulation results for the Floquet energies of the driven spin-$\frac{1}{2}$ chain system using algorithm Fauseweh-Zhu-1. Shown is the error according to Eq.\  \eqref{eq.error_algo1} as a function of layers in the Ansatz for increasing number of sites.} }
  \label{fig:figure4}
\end{figure}

{We choose an Ansatz $| \Psi_i (\mathbf{\Theta}) \rangle$ for the parametrized quantum circuit that has a layered structure, with each layer containing single qubit gates acting on all sites, which contain the variational parameters $\mathbf{\Theta}$, followed by entangling all qubits with CNOT gates. The details are shown in  Appendix \ref{appendix}. We  use the first algorithm to optimize the variational parameters for all quasi-energies. To quantify the convergence of the approach we define the error as the following $2$-norm
\begin{align}
\epsilon = \sqrt{ \sum\limits_i  \left(1 - |\left\langle \Psi_i (\mathbf{\Theta}) |  U_T \Psi_i (\mathbf{\Theta}) \right\rangle| \right)^2 }, \label{eq.error_algo1}
\end{align}
where $ \langle \Psi_i (\mathbf{\Theta}) |  U_T \Psi_i (\mathbf{\Theta}) \rangle$ measures the overlap between the initial state $| \Psi_i (\mathbf{\Theta}) \rangle$ of the approximate Floquet state with quasi-energie  $\epsilon_i$ and the time evolved state. The results are shown in Fig.\ \ref{fig:figure4} for up to $8$ sites in the chain. As expected, increasing the number of layers improves on the error of the variational approximation. Interestingly we observe the buildup of a plateau in the reachable accuracy with increasing chain length. Before the breakdown of this plateau the error reduces only slowly before it converges exponentially to zero with the number of layers.  This observation is similar to the recently observed scaling of variational quantum circuits representing ground states of systems at a quantum critical point \cite{BravoPrieto2020scalingof}. Our numerical analysis suggests that approximating Floquet eigenstates with variational quantum circuits is a hard problem similar to variationally approximating gapless quantum phases. This indeed makes sense, as a quantum critical point is driven by high-energy fluctuations, while Floquet systems are coupled across the whole (quasi)-energy spectrum. For ground states this behaviour breaks down once the quantum circuit is deep enough to resolve the finite size of the system, in which a finite spectral gap prevails. For Floquet systems a similar mechanism, in which the circuit can resolve the size of Floquet quasi-gaps, could provide an explanation for the breakdown of the plateau scaling.  }

\subsubsection{Algorithm Fauseweh-Zhu-2}

\begin{figure}[t]
  \centering
  \includegraphics[width=0.99\textwidth]{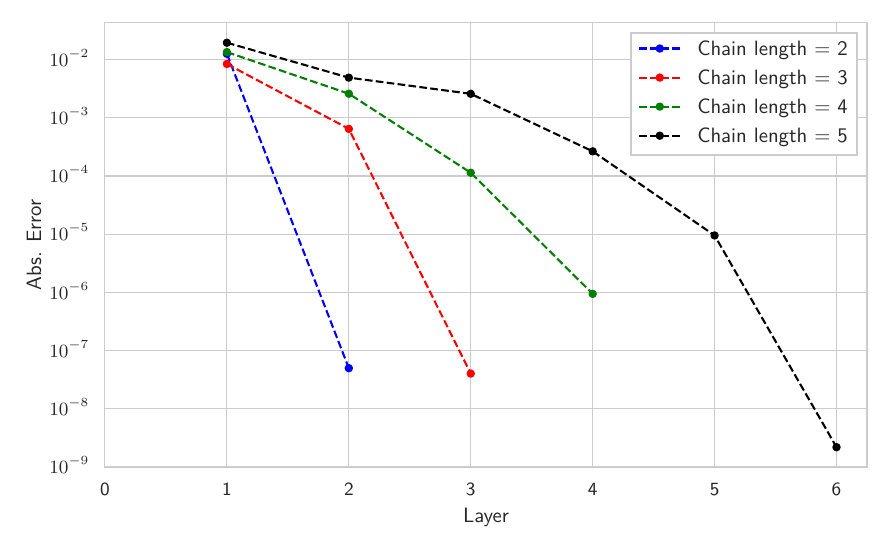}
  \caption{ {Simulation results for the Floquet energies of the driven spin-$\frac{1}{2}$ chain system using algorithm Fauseweh-Zhu-2. Shown is the error according to Eq.\  \eqref{eq.error_algo2} as a function of layers in the Ansatz for increasing number of sites.} }
  \label{fig:figure5}
\end{figure}

{The effective Hamiltonian of the driven spin-$\frac{1}{2}$ Heisenberg chain in the extended Hilbert space reads, }
{
\begin{align}
H_\mathrm{eff} =  - \frac{J}{4} \sum\limits_{i \in \mathbb{Z}_N} \sum\limits_{\alpha \in \lbrace x,y,z \rbrace} \sigma_{i, \alpha} \sigma_{i+1, \alpha}  \otimes \mathbb{1} + \sum\limits_{i \in \mathbb{Z}_N} \left[  A \sigma_{i,x} \otimes S_x + A \sigma_{i,y} \otimes S_y \right] + \Omega  \mathbb{1} \otimes S_z   \;.
\end{align}
}
{The Ansatz for the parameterized quantum circuit follows the definition in \eqref{eq.algo2_ansatz_general}. The generators are given in  Appendix \ref{appendix}. We define the error as the difference between the exact quasi-energy and the variational quasi-energy averaged over all Floquet states,  }
{
\begin{align}
\label{eq.error_algo2}
\epsilon = || \vec{E}_\text{exact} - \vec{E}_\text{var} || = \sqrt{ \sum\limits_i |E_{i, \text{exact}} - E_{i, \text{var}} |^2 }.
\end{align}}
{The results, shown in Fig.\ \ref{fig:figure5}, reveal a similar behavior as the first algorithm but with a slightly increased prefactor.  Thus the usage of the extended Hilbert space does not change the scaling of the representability of Floquet states with variational quantum circuits. Due to the increased simulation requirements coming with the qudit-qubit structure we only simulated spin chains with up to $5$ sites in this case.}

\section{Discussion}

In this paper we have presented two quantum algorithms, Fauseweh-Zhu-1 and Fauseweh-Zhu-2, to compute Floquet eigenstates and quasi-energies. Both algorithms use parametrized quantum circuits in combination with a quantum-classical hybrid approach to find solutions to the defining Floquet properties in time- and frequency-domain, respectively. While the precision of the first algorithm depends on the depth of the quantum circuit due to the IQPE application, the precision of the second algorithm mainly depends on the frequency truncation $j_\mathrm{max}$ and thereby on the width of the parametrized quantum circuit.
There are no fundamental hurdles in applying both algorithms on NISQ devices, with the advantage of complementary requirements in circuit depth and width for increasing system size. We have tested both algorithms on a quantum computer simulator for the linearly driven spin-$\frac{1}{2}$ problem and showed the principal applicability of our approach.
To investigate the performance of our algorithms for larger systems we simulated a circularly driven spin-$\frac{1}{2}$ chain with up to $8$ sites. We have demonstrated the feasibility of our variational approach and observed a scaling behavior in the accuracy of both approaches which was previously seen for ground states of quantum critical systems. Our work has uncovered a connection between variationally approximating ground state properties of critical systems and Floquet modes. Exploring the universality of this observation is left for future work. We have also simulated the effect of noise in Appendix \ref{appendix.b}, demonstrating the stability of our variational approach towards this perturbation.
As with other variational hybrid algorithms, the ansatz for the parametrized quantum circuit is fundamental for the success of the approach. In this context, the qudit-qubit structure of the second algorithm calls for novel schemes that have so far not been explored. It will be interesting to explore more complicated driving schemes with our approach and to test the performance on real devices using advanced error mitigation methods \cite{Fauseweh2021} in the near future. 

\section*{Acknowledgments}
We thank Andrew Sornborger, Zoe Holmes, Michael Epping, Tim Bode, Peter Schuhmacher, Elisabeth Lobe and Tobias Stollenwerk for useful discussions on related problems. This work was carried out under the auspices of  the  U.S.  Department  of  Energy  (DOE)  National Nuclear  Security  Administration  under  Contract  No. 89233218CNA000001. It  was supported by the LANL LDRD Program.

%\bibliographystyle{unsrtnat}
%\bibliography{literatur}

\onecolumn\newpage
\appendix

\section{Generators for variational Hamiltonian ansatz}
\label{appendix}

\subsection{Linearly driven spin-$\frac{1}{2}$}

To find an efficient variational ansatz for the Fauseweh-Zhu-2 algorithm we first separate the gates required for the $\mathcal{T}$ and the $\mathcal{R}$ part of the Hilbert space. For $j_\mathrm{max} = \pm 1$ the qudit space is $3$ dimensional. A general unitary matrix in SU($3$) can be parametrized using the Gell-Mann matrices, which span the corresponding Lie-Algebra. For our particular case we find, that a much more limited set is sufficient to find all Floquet eigenstates. For $\mathcal{R}$ a simple Ry gate is sufficient. 
Naturally for nonzero $A$ the Floquet eigenstates do not separate into the two Hilbert spaces anymore and we need an entangling gate to account for this.
We use the matrices
\begin{align}
S_4 = \begin{pmatrix}
0 & 0 &  0  \\
0 & 0 &1 \\
0 & 1& 0\\
\end{pmatrix} \quad S_5 = \begin{pmatrix}
0 & 1& 0  \\
1 & 0& 0 \\
0 & 0 & 0\\
\end{pmatrix} \quad S_x = \frac{1}{2} \left( S_5 + S_4 \right),
\end{align}
to define our variational Hamiltonian ansatz
\begin{align}
U_{\mathbf{\Theta}} = e^{i \Theta_1 \sigma_y} e^{i \Theta_2 S_4} e^{i \Theta_3 S_5} e^{i \Theta_4 S_x \sigma_x} e^{i \Theta_5 \sigma_y} e^{i \Theta_6 S_4} e^{i \Theta_7 S_5},
\end{align}
where $e^{i \Theta_4 S_x \sigma_x}$ is the aforementioned entangling gate. {For $j_\mathrm{max} = \pm 2$ we use a similar Ansatz by replacing the matrices $S_4$ and $S_5$ with the following generators of the SU$(5)$}
\begin{align}
S_{1} = \begin{pmatrix} 
        0& 1&  0&  0&  0\\
        1& 0&  0&  0&  0\\
        0& 0&  0&  0&  0\\
        0& 0&  0&  0&  0\\
        0& 0&  0&  0&  0
\end{pmatrix} &\quad
S_{2} = \begin{pmatrix} 
        0& -i&  0&  0&  0\\
        i& 0&  0&  0&  0\\
        0& 0&  0&  0&  0\\
        0& 0&  0&  0&  0\\
        0& 0&  0&  0&  0
\end{pmatrix} &\quad
S_{3} = \begin{pmatrix} 
        1& 0&  0&  0&  0\\
        0& -1&  0&  0&  0\\
        0& 0&  0&  0&  0\\
        0& 0&  0&  0&  0\\
        0& 0&  0&  0&  0
\end{pmatrix} \nonumber \\
S_{4} = \begin{pmatrix} 
        0& 0&  1&  0&  0\\
        0& 0&  0&  0&  0\\
        1& 0&  0&  0&  0\\
        0& 0&  0&  0&  0\\
        0& 0&  0&  0&  0
\end{pmatrix} & \quad
S_{5} = \begin{pmatrix}
        0& 0&  -i&  0&  0\\
        0& 0&  0&  0&  0\\
        i& 0&  0&  0&  0\\
        0& 0&  0&  0&  0\\
        0& 0&  0&  0&  0
\end{pmatrix} &\quad
S_{8} = \begin{pmatrix} 
        1& 0&  0&  0&  0\\
        0& 1&  0&  0&  0\\
        0& 0&  -2&  0&  0\\
        0& 0&  0&  0&  0\\
        0& 0&  0&  0&  0
\end{pmatrix} \nonumber \\
S_{9} = \begin{pmatrix} 
        0& 0&  0&  1&  0\\
        0& 0&  0&  0&  0\\
        0& 0&  0&  0&  0\\
        1& 0&  0&  0&  0\\
        0& 0&  0&  0&  0
\end{pmatrix} &\quad
S_{10} = \begin{pmatrix} 
        0& 0&  0&  -i&  0\\
        0& 0&  0&  0&  0\\
        0& 0&  0&  0&  0\\
        i& 0&  0&  0&  0\\
        0& 0&  0&  0&  0
\end{pmatrix} & \quad
S_{11} =  \begin{pmatrix} 
        0& 0&  0&  0&  1\\
        0& 0&  0&  0&  0\\
        0& 0&  0&  0&  0\\
        0& 0&  0&  0&  0\\
        1& 0&  0&  0&  0
\end{pmatrix} \nonumber \\
S_{12} = \begin{pmatrix} 
        0& 0&  0&  0&  -i\\
        0& 0&  0&  0&  0\\
        0& 0&  0&  0&  0\\
        0& 0&  0&  0&  0\\
        i& 0&  0&  0&  0
\end{pmatrix} &\quad
S_{24} = \begin{pmatrix} 
        -2& 0&  0&  0&  0\\
        0& -2&  0&  0&  0\\
        0& 0&  -2&  0&  0\\
        0& 0&  0&  3&  0\\
        0& 0&  0&  0&  3
\end{pmatrix} & .
\end{align}
{In this case, the matrix $S_x$ is given as,}
\begin{align}
S_{x} = \begin{pmatrix} 
        0& 1&  0&  0&  0\\
        1& 0&  1&  0&  0\\
        0& 1&  0&  1&  0\\
        0& 0&  1&  0&  1\\
        0& 0&  0&  1&  0
\end{pmatrix} &\quad .
\end{align}
{The part of the Ansatz acting only on the single qubit remains invariant. A single layer of this Ansatz is sufficient to produce the results shown in Fig. \ref{fig:figure3}.}

\subsection{Circularly driven spin-$\frac{1}{2}$ Heisenberg chain}

\begin{figure}[t]
  \centering
  \includegraphics[width=0.99\textwidth]{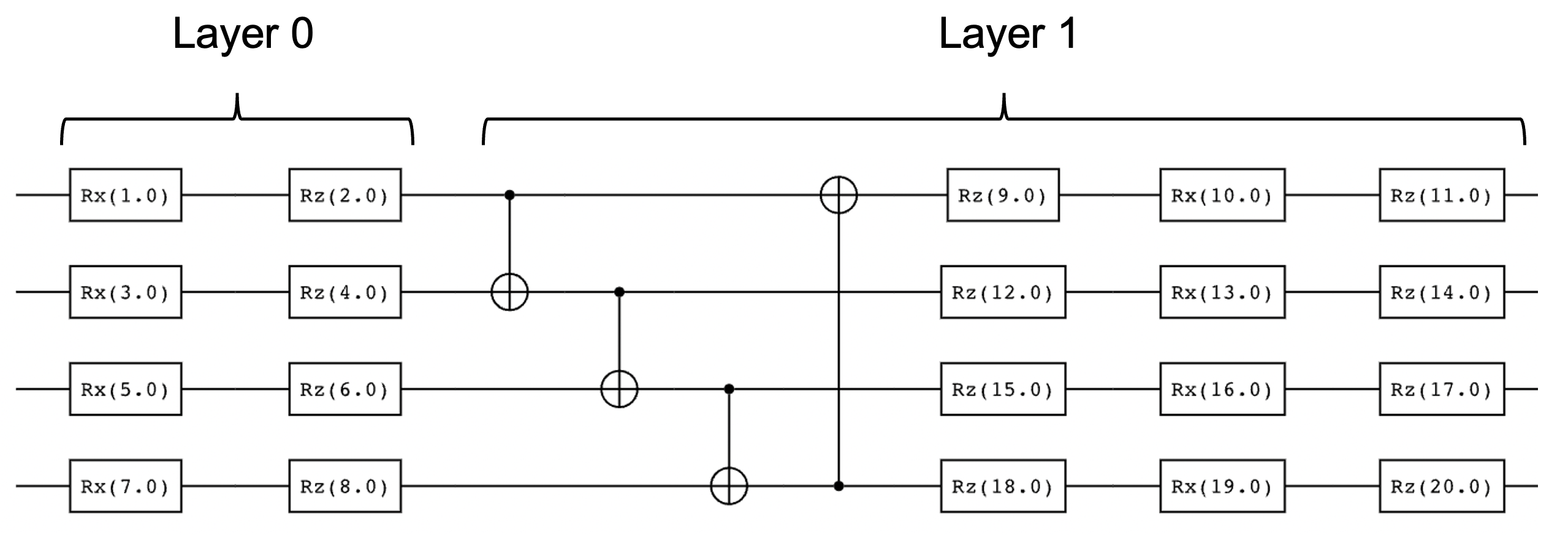}
  \caption{ {Variational circuit of the first algorithm for the driven spin-$\frac{1}{2}$ Heisenberg chain for $4$ sites and a single layer.} }
  \label{fig:figure7}
\end{figure}

{
For the driven Heisenberg spin chain we use a layered variational circuit that consists of an initial layer of single qubit gates, followed by a varying number of layers containing a linear chain of CNOT entangling gates and general single qubit rotation gates. A simple example for a $4$-site chain with a single layer is shown in Fig. \ref{fig:figure7}. For the second algorithm we use a layered ansatz of the following form,
\begin{align}
U_{\mathbf{\Theta}} &= \prod\limits_l U_{\mathbf{\Theta}, l} \\
U_{\mathbf{\Theta}, l} &= \prod_{i=1}^{N} e^{i \tilde\Theta \sigma_{i,z}} e^{i \tilde\Theta \sigma_{i,y}} e^{i \tilde\Theta \sigma_{i,z}} \prod_j e^{i \tilde\Theta \Gamma_j} \prod_{i=1}^{N}  e^{i \tilde\Theta \sigma_{i,x} \sigma_{i+1,x}}  \prod_{i=1}^{N} e^{i \tilde\Theta S_x \sigma_{i,x}} ,
\end{align}
}
{where $\tilde\Theta$ is an index function that assigns a unique parameter from the vector $\mathbf{\Theta}$ to the unitary operator and $ \Gamma_j$ are the Gell-Mann matrices \cite{PhysRev.125.1067}. The ansatz therefore consists of single qubit and qudit rotations followed by qubit-qubit entangling gates and finally qubit-qudit entangling gates.
}

\section{Noisy simulations}
\label{appendix.b}

\begin{figure}[t]
  \centering
  \includegraphics[width=0.99\textwidth]{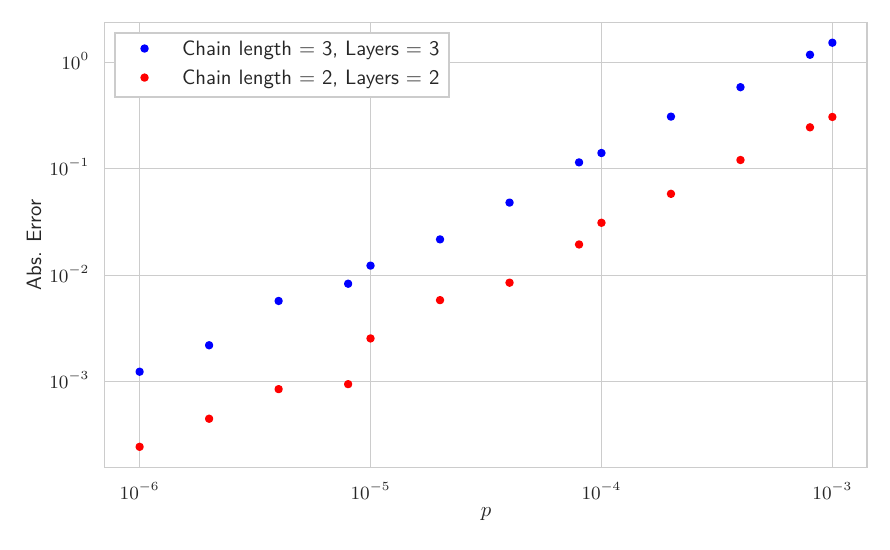}
  \caption{ {Results for the noisy simulation of the driven spin-$\frac{1}{2}$ chain system using algorithm Fauseweh-Zhu-2. Shown is the error according to Eq.\  \eqref{eq.error_algo2} as a function of the noise probability $p$ for a symmetric depolarization channel.} }
  \label{fig:figure6}
\end{figure}

{To evaluate the performance of our algorithms in the presence of noise we simulate the second algorithm in the presence of symmetric depolarising noise \cite{nielsen_chuang_2010} in the driven spin-$\frac{1}{2}$ Heisenberg chain. We employ the same variational Ansatz used in Fig.\ \ref{fig:figure5} for a chain length of $2$ and $3$ sites and measure the error given in Eq.\ \eqref{eq.error_algo2}. The noise channel is implemented into the simulation using the operator $ \Lambda_l$, modifying the resulting unitary evolution in Eq.\ \eqref{eq.algo2_ansatz_general} to }
{
\begin{align}
\tilde{U}_{\mathbf{\Theta}} = \prod\limits_{l} \exp( - i K_l \Theta_l) \Lambda_l.
\end{align}
}
{The noise operators act only on the qubit part of the system, neglecting decoherence in the qudit subspace as well as correlated noise between the qubits. We choose a symmetric depolarising noise, so that the noise operators are randomly chosen $\sigma_x$, $\sigma_y$ or $\sigma_z$ matrices acting on a random qubit $i$. We denote the probability of the noise operator deviating from the identity with $p$. We also take a finite shot noise into account by averaging over $10^5$ simulations. The results for the error are shown in Fig.\ \ref{fig:figure6}. For the low noise simulated here we see a  linear connection between the  probability of noise occurring with the error of the circuit independent from the system size. However the overall amplitude of the error indeed increases with the number of qubits. This behavior is similar to previous works investigating the effect of noise on variational quantum circuits \cite{PhysRevA.104.022403}, demonstrating the applicability of the approach for sufficiently low noise and the scaling behavior of the noise-induced error with the system size.} 

\end{document}